# Enhanced atomic layer etching of native aluminum oxide for ultraviolet optical applications

John Hennessy,[a*] Christopher S. Moore,[bc] Kunjithapatham Balasubramanian,[a] April D. Jewell,[a] Kevin France,[bc] Shouleh Nikzad[a]

[a] *Jet Propulsion Laboratory, California Institute of Technology, 4800 Oak Grove Dr., Pasadena, California 91109*
[b] *Center for Astrophysics and Space Astronomy, University of Colorado, Boulder, Colorado 80309*
[c] *Laboratory for Atmospheric and Space Physics, University of Colorado, Boulder, Colorado 80309*

[*]email: john.j.hennessy@jpl.nasa.gov

## Abstract
We report on the development and application of an atomic layer etching (ALE) procedure based on alternating exposures of trimethylaluminum and anhydrous hydrogen fluoride (HF) implemented to controllably etch aluminum oxide. Our ALE process utilizes the same chemistry previously demonstrated in the atomic layer deposition of aluminum fluoride thin films, and can therefore be exploited to remove the surface oxide from metallic aluminum and replace it with thin fluoride layers in order to improve the performance of ultraviolet aluminum mirrors. This ALE process is modified relative to existing methods through the use of a chamber conditioning film of lithium fluoride, which is shown to enhance the loss of fluorine surface species and results in conformal layer-by-layer etching of aluminum oxide films. Etch properties were explored over a temperature range of 225 to 300 °C with the $Al_2O_3$ etch rate increasing from 0.8 to 1.2 Å per ALE cycle at a fixed HF exposure of 60 ms per cycle. The effective etch rate has a dependence on the total HF exposure, but the process is shown to be scalable to large area substrates with a post-etch uniformity of better than 2% demonstrated on 125 mm diameter wafers. The efficacy of the ALE process in reducing interfacial native aluminum oxide on evaporated aluminum mirrors is demonstrated with characterization by x-ray photoelectron spectroscopy and measurements of ultraviolet reflectance at wavelengths down to 120 nm.

## Introduction

Aluminum possesses a higher ultraviolet (UV) reflectance than other metals like gold and silver due to material properties such as a higher plasma frequency and the absence of strong interband transitions at UV wavelengths. These properties of Al are employed in classical UV optical components like mirrors,[1,2] reflective diffraction gratings,[3,4] and multilayer UV bandpass filters.[5,6] Al is also gaining interest as a plasmonic material for UV applications. The same material properties lead to a wider spectral tunability range for plasmonic effects in Al,[7] and nanostructured Al is being considered for applications like UV resonant biochemical sensors,[8] UV nanoantennas,[9,10] and UV transmission filters.[11,12]

The difficulty in working with Al, particularly for UV applications, is associated with its tendency to rapidly oxidize when exposed to air. This makes the bulk properties of Al notoriously challenging to control as a function of deposition conditions, and the presence of the surface oxide itself can significantly alter the performance of both classical[1,2] and plasmonic devices[7,13] due to strong optical absorption at wavelengths shorter than 250 nm.

The allure of the predicted performance of ideal, non-oxidized Al in the far UV (FUV, 90 < $\lambda$ < 200 nm) has led to several proposals for the on-orbit fabrication of Al mirrors where the vacuum of space may prevent or slow oxidation.[14,15] Conventional fabrication techniques for Al UV mirrors prevent this oxidation by depositing a protective layer immediately following Al evaporation. These protective layers are typically metal fluoride materials that are transparent at UV wavelengths. For example, a large number of astronomical UV mirrors, *e.g.*, the Hubble Space Telescope optical telescope assembly, are fabricated with evaporated Al followed by 25 nm of $MgF_2$.[16] Alternate materials such as $LiF$[17] and $AlF_3$[18] can also offer good performance with varying short wavelength cutoffs in the FUV related to the optical bandgap of the protective coating material, and with tradeoffs generally associated with the environmental stability of the protective coating.[19] The advantage of depositing these layers with conventional physical vapor deposition (PVD) methods is that this can be performed sequentially with Al PVD, but the drawback is related to the minimum thickness required by PVD methods to achieve good uniformity and pinhole free films, particularly as these layers are often deposited at high rates.[17,19,20]

The protective coating can limit the reflective performance of the mirror at FUV wavelengths due to optical absorption, because loss becomes proportional with the thickness of the absorbing overcoat. There is therefore considerable interest in producing protected Al mirrors with coatings that are as thin as possible to minimize absorption losses, while still thick enough to prevent subsequent oxidation.[2] Atomic layer deposition (ALD) has been pursued toward this end due to its strength in forming dense, pinhole free films. Previous work in our lab has demonstrated ALD processes for candidate fluoride protective coating materials,[4,21] as well as the fabrication of prototype ALD-protected Al mirrors.[2,22,23]

High performance metallic Al has not been demonstrated with ALD methods, so a combination of evaporated Al and ALD overcoat has been required to fabricate these components. Combined or shared vacuum chamber systems are challenging due to the dual requirement of properly isolating ALD chemistry from contaminating evaporation sources and chambers, and the low temperature and ultra-high vacuum (UHV) needed for Al evaporation. A vacuum transfer approach may alleviate some of these issues, but is challenging for larger sample sizes and becomes limited by issues associated with the time required for substrate transfer and heating/cooling. The reflectance of Al has been observed to degrade even in UHV conditions,[1] so any delay prior to encapsulation is likely to be detrimental as even several angstroms of interfacial oxidation can limit the reflective performance at FUV wavelengths.[2]

It is therefore desirable, particularly for ALD encapsulation, to explore methods to remove the surface oxide *in situ* prior to the deposition of the fluoride protective layer.

Traditional microfabrication process methods like chemical sputtering[24] or ion etching[25] can be used to strip this oxide, but leave the metallic surface exposed and susceptible to reoxidation. Additionally, for applications employing structured Al layers, it is not practical to perform lithographic patterning and simultaneously prevent this oxidation from occurring. Applications requiring patterning and subsequent electrical contact to Al layers can benefit from the capability to strip a native oxide and immediately deposit ALD dielectric or metallic materials. For example, the controlled removal of Al surface oxide could improve the subsequent nucleation of additional deposited metals, and reduce the specific contact resistance of multi-layer metal stacks for integrated circuit applications.[25]

Atomic layer etching (ALE) *via* alternating surface reactions is an appealing option for this task, as the conformal and precisely controlled removal of oxide from the Al surface can easily be combined with ALD methods for fluoride deposition in the same vacuum chamber. ALE methods for the removal of $Al_2O_3$ have been demonstrated through the chemisorption of $BCl_3$, and the subsequent neutral Ar beam bombardment of the surface.[26] More recently, several alternate thermal approaches have also been demonstrated that do not rely on exposure to energetic species which may damage the sensitive Al surface. One example utilizes alternating exposure to hydrogen fluoride (HF) and tin(II) acetylacetonate,[27] and another implements a similar approach with trimethylaluminum (TMA) and HF.[28]

In this work, we investigate the use of a modified version of this TMA-based thermal ALE approach and demonstrate the efficacy of this method in improving the UV performance of evaporated Al thin films. The reduction in surface native $Al_2O_3$ is characterized by near ultraviolet (NUV, $200 < \lambda < 400$ nm) and FUV reflectance measurements, and the prospects for applications related to future astronomical mirrors are discussed.

**Experimental Section**

ALD and ALE experiments were conducted in a showerhead-style reactor (Oxford OpAL) using alternating pulses of TMA and anhydrous HF; or alternatively with TMA and water vapor for the deposition of $Al_2O_3$. Anhydrous HF [Matheson Tri Gas, 4N5] was stored at room temperature and delivered through a single stage regulator and needle valve. The HF cylinder was enclosed within the separately ventilated precursor cabinet and pulsed into the reaction chamber using standard ALD diaphragm valves. The ALD purge gas was Ar [Matheson Tri Gas, 6N], and was delivered through a mass flow controller at 70 sccm. The configuration of our showerhead-style system involves a heated substrate table within a larger heated chamber volume. The table temperature is variable in order to set the substrate temperature, and the remainder of the internal chamber is held at 150 °C in order to prevent precursor condensation on internal surfaces and protect the survivability of the outer o-rings responsible for isolating the precursor delivery paths and maintaining vacuum conditions.

Aluminum thin films were deposited by electron beam evaporation in a load-locked system (AJA International), with the deposition chamber reaching a typical base pressure of less than 2 x $10^{-9}$ Torr. Al was deposited to target thickness of 60 nm at a deposition rate of 1.5 nm/sec; film thickness and evaporation rate were monitored by quartz crystal microbalance.

The surface topology of reflective Al samples was imaged with a Zygo ZeMapper white light interferometer at lengths scales of 100 µm; images were collected with a magnification corresponding to sampling at 93 nm per pixel. X-ray photoelectron spectroscopy (XPS) analysis was performed at the Beckman Institute at the California Institute of Technology. Spectra were collected on a Surface Science M-Probe ESCA equipped with an Al Kα monochromatic x-ray source at 1486.6 eV and chamber pressure <8×$10^{-9}$ Torr.

NUV optical characterization was performed on deposited and etched $AlF_3$ and $Al_2O_3$ thin films. Film thickness and refractive index were measured by *ex situ* spectroscopic ellipsometry on a Horiba Uvisel 2 phase-modulated ellipsometer. ALD-protected Al samples and unprotected control samples where characterized for near-normal incidence NUV reflectance using a Perkin Elmer Lambda 1050 spectrophotometer at near atmospheric pressure in a nitrogen-purged environment. FUV reflectance was measured on an Acton VM 502 monochromator operating under vacuum with a deuterium lamp over an effective wavelength range of 120–180 nm.

## Results and Discussion
*ALE Process Development*

The ALE procedure described in this work is based on alternating exposures of TMA and anhydrous HF. The same chemistry has been used in our lab for the deposition of $AlF_3$ thin films for UV optical applications.[21] In this previous work, we noted that the deposition rate of these films steadily decreased with increasing substrate temperature, and that $AlF_3$ film etching was possible at higher temperatures. Here we use the same chemistry to etch $Al_2O_3$ films, the reaction proceeds by converting the surface oxide to fluoride during exposure to HF, followed by $AlF_3$ etching in the same manner as noted in our previous work.

This method is similar to recent reports that note that TMA is effective as a metal etch precursor in combination with exposure to HF sourced from a solution of HF-pyridine.[28] This technique has been extended to variety of oxide and nitride materials, and has several potential applications related to self-limiting selective etching or the development of conformal high-aspect ratio etch processes.[29,30] Our approach is similar to this alternate report with some key differences. First, rather than a solution of HF-pyridine, this work uses anhydrous HF as the fluorine source. HF-pyridine has been utilized in previous reports on ALD metal fluoride processes.[31,32] Mass spectroscopy studies combined with the high partial pressure of HF over the solution suggested that pyridine does not participate in the reaction process.[31] It is expected that the same comparison is true in this ALE experiment, and that the use of HF-pyridine is a functional analog to the use of anhydrous HF, although the delivered HF dose may be different. At the temperatures and pressures explored in our experiment, the surface fluorination of deposited $Al_2O_3$ films is not easily saturating within typical ALD dosing regimes and leads to considerable tunability of the effective etch rate.

A second difference is related to our use of a chamber conditioning step. Lee *et al.* have characterized the etching of $Al_2O_3$ as a combination of surface fluorination during HF exposure and eventual mass loss *via* a transmetalation reaction that occurs during TMA exposure. Our observations indicate that chamber background conditioning plays an important role in both the etch rate and spatial uniformity of etched $Al_2O_3$ films. For some background conditions ALE can be entirely suppressed, even at the maximum explored temperature of 300 °C. These effects suggest that a secondary reaction mechanism is responsible for mass loss during TMA exposure.

Background conditioning with ALD lithium fluoride films was used for all ALE experiments performed in this work. Lithium fluoride was deposited at 150 °C onto to the substrate table and interior chamber at thicknesses between 5-15 nm using an ALD process involving alternating exposures of lithium bis(trimethylsilyl)amide and anhydrous HF.[32,33] The absolute thickness of the conditioning film does not appear to alter observed $Al_2O_3$ etch rates, suggesting that the surface species loss that occurs during TMA exposure remains self-limiting despite the secondary interaction. Furthermore, XPS analysis of etched $Al_2O_3$ films does not indicate the presence of a detectable amount of lithium incorporation, suggesting that lithium exchange is not a major component of the etch mechanism. This hypothesis also supported by the observation that the etch procedure is self-sustaining once the conditioning LiF film is present. For example, continuous etching of nearly 500 nm of $Al_2O_3$ was performed following deposition of only 15 nm of LiF with no measured variation in effective etch rate. We propose a fluorine exchange during TMA exposure that results in a byproduct that either directly etches the surface fluoride, or that promotes the tranmetalation of adborbed $Al_xF_y(CH_3)_z$ species. The strong Lewis acid nature of TMA combined with the typical basic and ionic properties of LiF may allow this reaction to proceed.[34] The mechanism of the interaction between TMA and LiF will be the subject of additional study.

Process development and verification was performed on small area silicon coupons approximately 1 $cm^2$ cleaved from larger wafers in order to evaluate the efficacy of this technique for the modification of oxidized Al. Effective oxide etch rates were determined from ellipsometric measurements on pre-deposited ALD $Al_2O_3$ films as a proxy for the native aluminum oxide. Unless otherwise indicated, the nominal single dose ALE cycle in this experiment consisted of a single TMA exposure of 45 or 60 ms, followed by 10 s evacuation; and then a single HF exposure of 60 ms, followed by 45 s evacuation. The process was performed under a continuous inert purge flow of Ar. Purge times were determined by observation of saturation in the measured $Al_2O_3$ etch rate, for significantly shorter purge times the measured etch rate is reduced due to parasitic gas phase reactions that compete with the etch mechanism.

Figure 1 shows the effective etch rate of $Al_2O_3$ as a function of substrate temperature compared to a similar characterization in previous reports.[28] In this example, the TMA exposure was fixed at 45 ms pulse per cycle, and HF exposure was varied to highlight the significant dependence of etch rate on surface fluorination. For TMA pulse times greater than or equal to 30 ms there was little impact on the measured etch rate for small area samples. The observed increase in measured etch rate over previous reports is attributed to the more complete removal of the surface fluoride layer when utilizing the co-reaction with LiF.

The nominal HF exposure of 60 ms was chosen based on the typical exposure times required for the deposition of ALD AlF$_3$ at lower substrate temperatures.[21] At this pulse time the pressure transient introduced during the HF pulse is approximately 500 mTorr greater than idle base pressure of 200 mTorr, with an inert Ar purge flow of 70 sccm. At minimum valve pulse times of 15 ms this pressure transient can be reduced to approximately 200 mTorr over base pressure, but a saturation in measured etch rate for these shorter exposures was also not encountered. A single 15 ms HF pulse per cycle yielded an effective etch rate of 1.0 Å/cycle at 300 °C as shown in Figure 1. HF pulse times of 90 ms per cycle result in pressure transients of 800 mTorr over base pressure. Longer single-dose pulse times were not explored due to a process interlock that limits extended exposure to chamber pressures above 1 Torr in order to prevent accidental venting during HF processing.

As noted above, increasing HF exposure can lead to increased etch rates. This is emphasized in Figure 2, which shows the effect of replacing a single HF pulse with multiple HF pulses per cycle at both 300 and 225 °C. In this case, the TMA exposure was a single 60 ms pulse per cycle; individual 60 ms HF pulses were separated by a 1 s purge, and ultimately followed by a standard 45 s purge prior to the next cycle. At higher temperatures, the observed rate increases significantly with the number of HF pulse per cycle. No obvious saturation in the measured etch was observed for multiple HF pulses per cycle (up to 30 doses) which resulted in etch rates greater than 5 Å/cycle. At lower temperatures, a more apparent saturation is observed when more than ten 60 ms HF pulses per cycle are used. This trend implies a thermal component to the fluorination of the deposited oxide films.

Although saturation in the etch rate for multiple HF pulses per cycle is observed at lower temperatures, the practicality of adding this complexity to the ALE cycle was deemed undesirable. Additionally, the etch uniformity over larger substrate areas appeared to be degraded in our system for multiple HF pulses per cycle. Continued exposure to HF can presumably enforce both saturation and uniformity, although it would require increased HF consumption and possible modifications to the gas delivery and purging mechanisms, potentially using an additional mass flow controller, or by operating the HF exposure in a static configuration (no pumping). This approach will lead to higher etch rates, especially at higher temperatures near 300 °C where no obvious saturation was observed.

Despite the observation of a HF-dose dependence, the linearity of the single-dose etch procedure remains good as indicated in Figure 3. Although not limited to immediate surface sites, the fluorination during HF exposure is presumably bound by the diffusivity of reactive fluorine species into the bulk oxide as a function of temperature and pressure. This phenomenon may allow for some effective averaging of variations in HF surface concentration during the 60 ms exposure and subsequent purging that results in the observed linearity and the large area uniformity discussed below.

Exposure to longer or multiple TMA pulses per cycle does not increase the measured etch rate over the single dose etch cycle on small area samples. However, because the TMA interaction is driven by secondary reactions with the conditioning LiF deposited on the heated substrate table, a loading effect becomes apparent for larger area substrates that cover more of the available conditioning surface area. Figure 4a shows the measured etch uniformity of 125 mm silicon wafers pre-coated with ALD $Al_2O_3$ and exposed to 100 cycles of the single dose ALE cycle described above. This sample size is approximately 50% of the entire substrate table diameter. A clear increase in the amount of film etched is evident around the periphery of the wafer in closest proximity to the conditioning area. In the central portion of the wafer, the etch rate is reduced relative to that observed for small sample sizes. The etch rate and uniformity can be improved for large diameter samples by increasing the TMA dose as indicated in Figure 4b. In this case, four individual doses of TMA were delivered per etch cycle, with each dose separated by a 1 s purge, and the standard 10 s purge following all four. For each wafer in Figure 4, the starting thickness uniformity of the ALD $Al_2O_3$ was measured as 1.3% (± 1 σ). For the single TMA pulse per ALE cycle, the uniformity after 100 etch cycles was 10.0%, this was improved to 1.8% for the sample receiving four TMA pulses per cycle. Additionally, the measured etch rate, even in the central portion of the wafer, is approaching the small sample limit of 1.2 Å/cycle at 300 °C.

These observations serve as confirmation that secondary reactions are a dominant component of the etch mechanism, and that surface species loss (film thickness reduction) occurs during the TMA exposure. The loading effect has obvious implications on the efficacy of this conditioning approach for certain reactor designs where heated surface area may not be available. We expect an approach like this is still viable in showerhead-style systems with the introduction of a conditioned and heated showerhead or the use of a conditioned hot-wire array. In a cross-flow style ALD system a conditioned chamber ceiling or upstream reaction volume may be similarly effective.

*Application to Aluminum*

The application of this ALE method to the surface preparation of Al is motivated by the observation that the same process chemistry of TMA and HF is capable of etching $Al_2O_3$ as well as depositing $AlF_3$, which is effective as a protective layer with improved transparency in the UV. By changing the substrate temperature, it is possible to switch from the etch mode described above to a deposition mode where $AlF_3$ deposition dominates the reaction. Because the etch procedure relies on surface fluorination it is expected that some protection against reoxidation can also be provided. The transition from ALE to ALD is performed here by continuous alternating chemical exposure while the temperature is ramped, as opposed to pausing the dosing process to wait for the substrate temperature to change.

Initial tests were performed to explore the role of the starting etch temperature on the final reflectivity of evaporated Al films. Several Al samples were prepared and allowed to oxidize over a time scale of several days. Previous studies indicate that this time is sufficient to reach greater than 90% of the terminal room temperature native oxide thickness.[2] Etched samples were then processed with 50 single dose ALE cycles at a starting substrate temperature ($T_{SS}$) of 225, 250, or 300 °C before cooling to 150 °C while continuing to dose for 100 cycles (same pulse/purge times). This leads to a variable amount of etching and deposition, as the time required for the temperature change depends on the initial setpoint. For $T_{SS}$=300 °C the time required to reach 200 °C was 30 minutes, versus only 10 minutes for $T_{SS}$=225 °C. At a substrate temperature of 200 °C, deposition clearly dominates over etching; the additional reduction to 150 °C increases the deposition rate[21] and ensures proper encapsulation.

Figure 5 shows the resulting reflectance of etched samples compared to samples receiving only $AlF_3$ deposition at 150 °C (*i.e.*, no ALE cycles). Consideration of the ~1 min cycle time, the time required for deposition, and ellipsometric characterization of silicon witness samples suggests that the etched samples were encapsulated with 6–9 nm $AlF_3$, increasing with lower starting temperature. Samples etched at a higher starting temperature exhibit a significant degradation in reflectance compared to the control samples. As $T_{SS}$ is reduced, the measured reflectance is significantly increased, and for the lowest $T_{SS}$ of 225 °C, the measured reflectance begins to surpass that of the control samples. At $\lambda$~200 nm this sample also provides improved performance over a non-etched bare Al sample.

The degradation of reflectance for films etched at a higher starting temperature is the result of significant roughening of the metallic surface. Figure 6 shows interference micrographs of the etched samples measured in Figure 5 in comparison to a control sample. For $T_{SS}$=300 °C (sample a), aggressive pitting is observed with micron-scale holes greater than 10 nm in depth. Roughening is significantly reduced for samples with lower starting temperatures; some surface damage is still evident for $T_{SS}$=225 °C (sample c), but only at depths of ~1 nm. This damage is likely associated with the rapid fluorination of the underlying Al, potentially along grain boundaries, which may then etch more rapidly than regions where some oxide remains. This fluorination is presumed to have a strong thermal component similar to the $Al_2O_3$ etch dependence presented in Figure 2, which explains the large difference in surface damage.

Despite this surface damage, the impact of a reduction in interfacial oxide is readily apparent when comparing the FUV reflectance of ALD protected mirrors fabricated with and without ALE exposure. Figure 7 shows the near-normal incidence reflectance of the sample c and d mirrors over the FUV wavelength range of 120 to 180 nm. The similarity of the reflectance of these samples at $\lambda$ > 300 nm (Figure 5b), where losses from the both the ALD $AlF_3$ and the native Al oxide are negligible, indicates an equivalent total optical thickness of the combined dielectric layers. Therefore, the significant improvement in FUV reflectance can be attributed to a reduction in interfacial oxide, which is much more strongly absorbing in this wavelength range than the $AlF_3$ coating. Surface roughness can contribute significantly to the optical properties of Al,[35,36] particularly at FUV wavelengths,[37] but the larger spatial scale pitting observed here does not appear to severely impact the measured FUV performance.

In order to explore minimization of this damage, an additional optimization was investigated by fixing $T_{SS}$ at 225 °C and modifying the initial number of ALE cycles before starting the temperature ramp. In this case, the number of deposition cycles was reduced to 100 to explore reduced protective coating thicknesses, and the delivered HF dose was reduced to 30 ms during the deposition period to minimize over-etching. The resulting measured NUV reflectance is shown in Figure 8. Again, over-etching causes a degradation in reflectance but with less severity than higher temperature samples for the same effective number of ALE cycles. In other words, at an equivalent expected oxide removal rate (*e.g.*, predicted from Figure 1), the observed damage to Al is more severe at higher temperatures where the fluorination is predicted to be more aggressive.

With a reduced number of ALD cycles, these samples now have a protective coating thickness of only 3–4 nm, which is on the order of the expected thickness of the native oxide present prior to etching.[2,7,25] An optimal number of ALE cycles is apparent between 25 and 50, this is approximately consistent with this expected native oxide thickness and the expected oxide etch rate of 0.77 Å/cycle at 225 °C. These samples appear to exceed the reflectivity of bare oxidized Al at all measured NUV wavelengths, confirming a reduction in the amount of absorbing interfacial oxide, but also suggesting the cumulative optical thickness of the final protective layer is less than the starting native oxide. For samples that received more than 50 etch cycles, the reflectance degrades in a manner similar to the trend seen in Figure 5a. It can be seen that the degradation is less severe for 100 etch cycles at 225 °C (sample g) compared to 50 etch cycles at 250 °C (sample b), despite the similar expected amount of oxide etched based on the predicted etch rates plotted in Figure 1. This again suggests a strong thermal component the fluorination rate once the metallic Al is encountered, and that some selectivity is achieved by reducing the etch temperature.

The improved selectivity results in reduced surface damage to Al as indicated in Figure 9. The occurrence of surface pitting is eliminated in sample e, receiving 25 ALE cycles, and only barely perceptible in sample f, which received 50 ALE cycles. Compared to the earlier surface analysis of sample c (Figure 6), the reduction in HF dose as the temperature ramp is initiated appears to be effective at improving the measured surface roughness. Both samples surpass the non-etched control sample in NUV reflectance.

X-ray photoelectron spectroscopy was performed on ALE samples in order to confirm a chemical reduction in surface oxide. A comparison of the Al 2p spectra for etched sample f and a co-evaporated bare Al control sample is presented in Figure 10. Peak deconvolution was performed by fitting asymmetric line shapes to the metallic component near 73 eV and symmetric Gaussian-Lorentzian shapes to fit potential oxide and fluoride signals near 75 and 77 eV, respectively.[38] A small shift in relative binding energy is noted, as well as a slight distortion in the metallic component for sample f relative to the non-etched sample; the background-subtracted intensity plotted in Figure 10 is otherwise not normalized. The energy shift and distortion may be the result of variations in the surface charging for the ALD $AlF_3$ terminated surface relative to native oxide, or possibly a signature of the etch damage, which can alter the measured spectra. Nevertheless, a clear reduction in the oxide component is apparent for the etched sample capped with $AlF_3$, which is consistent with the improved reflectance.

## Conclusions

We have demonstrated a thermal ALE method capable of etching $Al_2O_3$ thin films with an enhanced etch rate and improved low temperature performance when compared to similar methods. This process was implemented to improve the UV performance of Al, and we expect it will have applications in both conventional Al UV optical devices as well non-planar Al surfaces such as Al nanostructured material used for UV plasmonic devices.

It is expected that additional improvements to this procedure are possible with further optimization of the starting etch temperature and the precise number of ALE cycles prior to encapsulation of the underlying Al. The enhancement in $Al_2O_3$ etch rate afforded by the intermediate reaction with LiF may allow for additional reductions in $T_{SS}$ in the intermediate temperature range where ALE of $Al_2O_3$ dominates over ALD of $AlF_3$. It may also be possible to tune this transition temperature by altering the delivered precursor dose or the working chamber pressure. Minimization of $T_{SS}$ appears to improve the ALE selectivity over metallic Al reducing the observation of associated damage. Adjusting the TMA and HF dosing parameters should also allow for additional improvements in the efficacy of this combined ALE/ALD approach, particularly as it relates to the fabrication of Al mirrors for future space astrophysics applications. Future NASA astrophysics missions will require improvements in reflectivity performance at wavelengths as close to 90 nm as possible.[39,40]

The demonstrated ALE technique will also have general applications as a conformal thin film removal method that can operate with high selectivity to materials not susceptible to same fluorination and volatilization as $Al_2O_3$. For example, the etching of $HfO_2$ and $AlF_3$ is readily observed with this LiF-enhanced ALE approach, but with little to no etching observed for Si, $SiO_2$, and $MgF_2$ among many other combinations. Process selectivity and materials limitations will also continue to be subject of continuing work.

## Acknowledgments

This research was performed at the Jet Propulsion Laboratory, California Institute of Technology, under a contract with the National Aeronautics and Space Administration. Support for C. S. Moore is provided through NASA Space Technology Research Fellowship (NSTRF) Program Grant # NNX13AL35H.

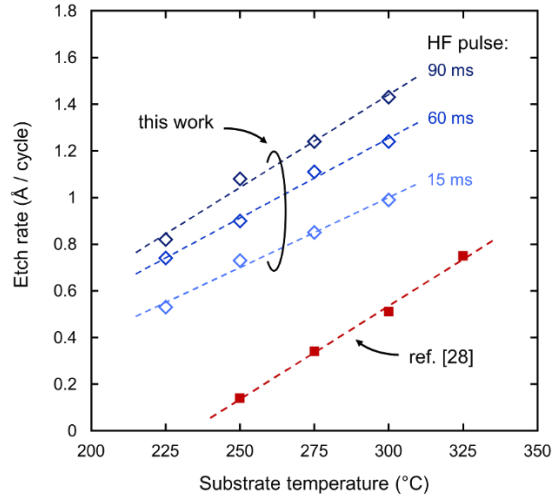

Figure 1. Observed etch rate of ALD $Al_2O_3$ as a function of substrate temperature for a single TMA pulse of 45 ms and variable HF pulse time in comparison to a similar process.[28] Samples were etched in a chamber previously passivated with LiF.

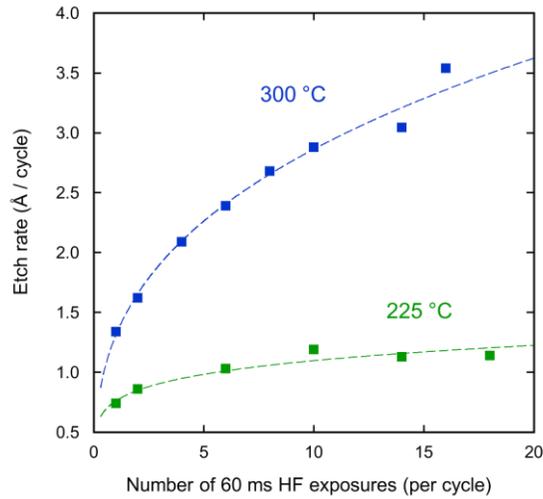

Figure 2. The measured etch rate for 100 ALE cycles incorporating multiple 60 ms HF pulses per cycle at substrate temperatures of 225 and 300 °C, the dashed lines indicate an approximate power law fit to the measured etch rate.

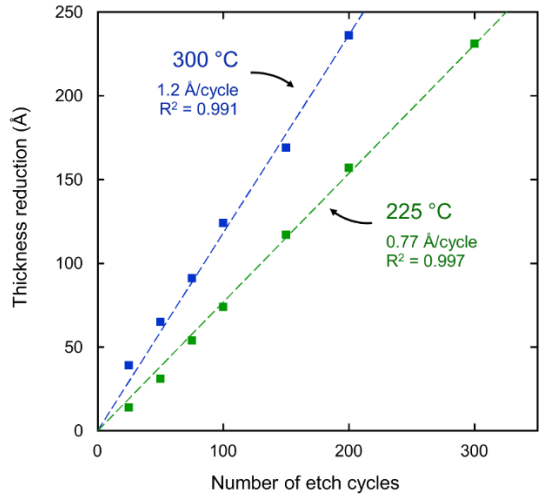

Figure 3. The linearity of the measured Al$_2$O$_3$ etch rate at 225 and 300 °C, for single 60 ms pulses of both TMA and HF.

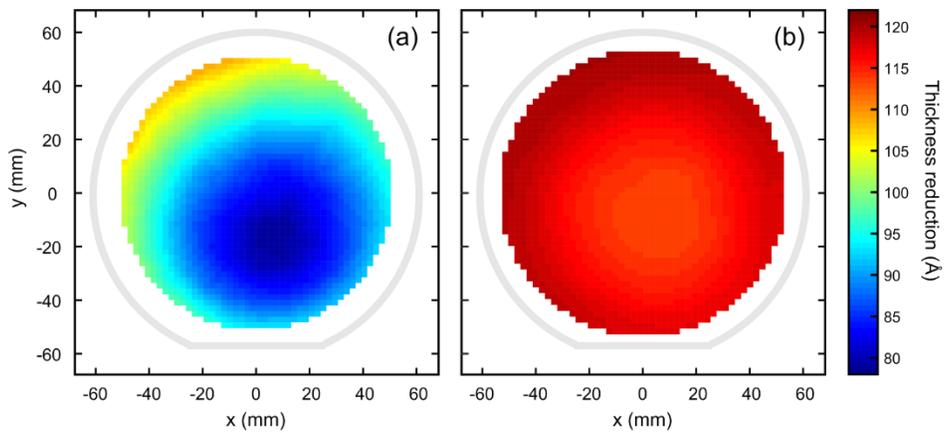

Figure 4. The measured spatial variation of etching uniformity over 125 mm silicon wafers for 100 ALE cycles including a single 60 ms HF pulse and (a) one 60 ms TMA pulse or (b) four 60 ms TMA pulses per cycle.

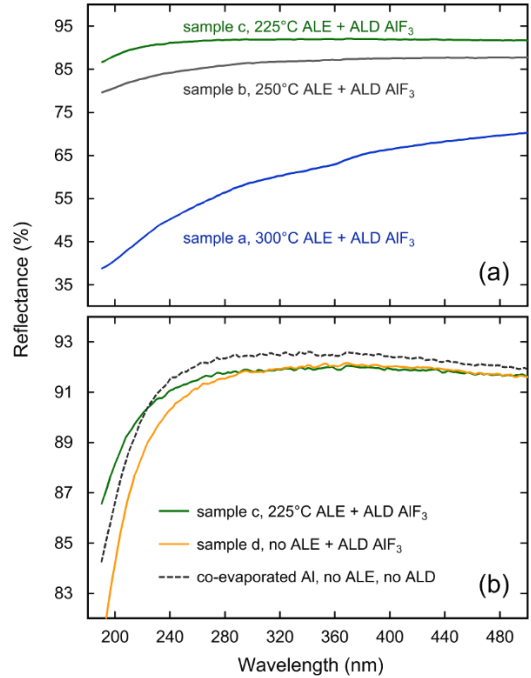

Figure 5. (a) The measured NUV reflectance of evaporated aluminum exposed to 50 ALE cycles and 200 ALD cycles for different starting substrate temperatures. (b) Non-etched sample d with a similar protective $AlF_3$ coating is shown for comparison along with a co-evaporated sample of bare aluminum.

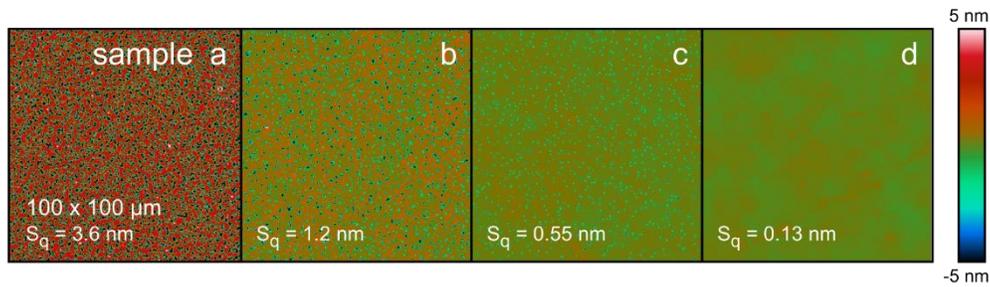

Figure 6. Interference micrographs of the Al surface height after exposure to 50 ALE cycles and 200 ALD cycles for different starting substrate temperatures. Sample labels correspond to the measured reflectances shown in Figure 5 (samples a-c), compared to the co-evaporated sample with no etch exposure and a similar ALD $AlF_3$ coating (sample d).

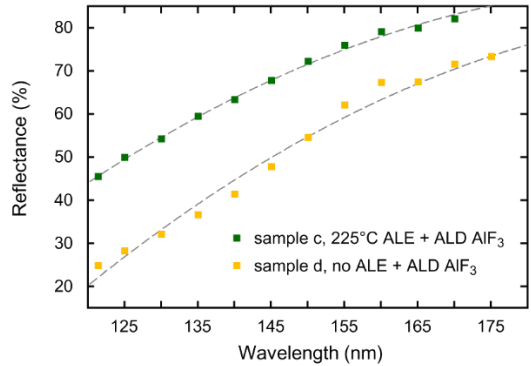

Figure 7. The measured FUV reflectance of evaporated Al exposed to 50 etch cycles at 225 °C and capped with 8–9 nm of ALD AlF$_3$ versus co-evaporated Al receiving no etch and capped with 6 nm ALD AlF$_3$ to yield a comparable total thickness when including the native Al oxide (samples c and d in Figure 5).

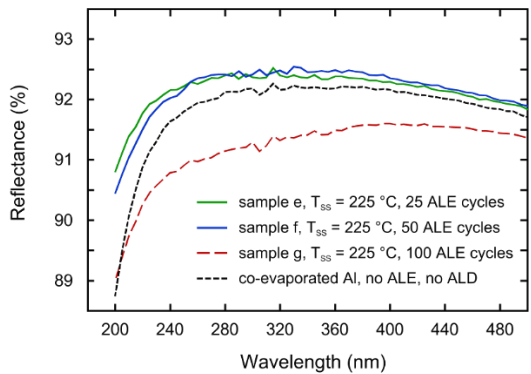

Figure 8. The measured NUV reflectance of evaporated aluminum exposed to a variable number of ALE cycles at 225 °C and 100 ALD cycles compared to a co-evaporated sample of non-etched Al.

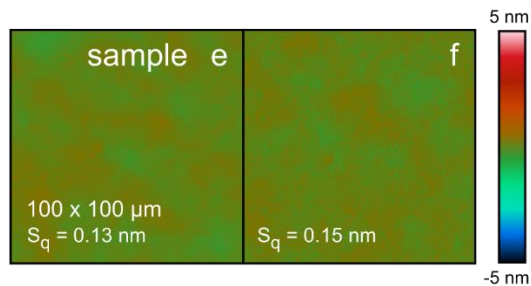

Figure 9. Interference micrographs of the Al surface height after exposure to 25 (sample e) or 50 ALE cycles (sample f) and 100 ALD cycles for $T_{SS}$ = 225 °C. The corresponding NUV reflectance of each sample is shown in Figure 8.

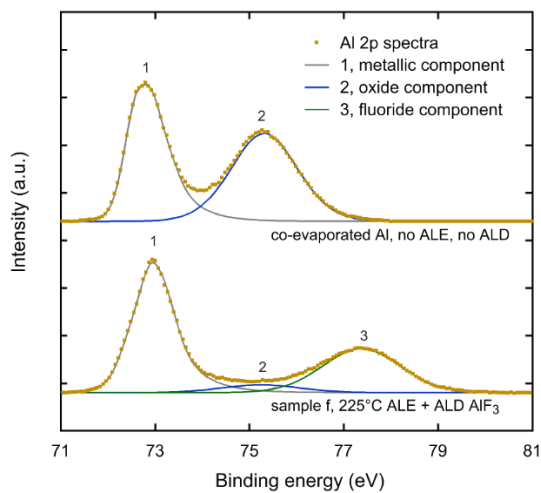

Figure 10. High-resolution XPS spectra and the background-adjusted components of the Al 2p level for evaporated Al with native oxide compared to co-evaporated films exposed to 50 etch cycles at 225 °C and 100 deposition cycles while the temperature was ramped to 150 °C. For clarity the binding energy (BE) of the etched sample f was shifted toward lower BE by 0.3 eV.